\documentstyle[12pt]{article}
\begin{document}
\newcommand{\suml}{\sum\limits}
\newcommand{\dvolumes}{\prod\limits_{k=1}^3d\Omega_{\p'_k}}
 \newcommand{\dvol}{\prod\limits_{k=1}^N(d\Omega_{\p'_k}d\Omega_{\p_k})}
 \newcommand{\p}{{\bf p}}
 \newcommand{\k}{{\bf k}}
 \newcommand{\K}{{\bf K}}
 \newcommand{\0}{{\bf 0}}
 \newcommand{\Q}{{\bf Q}}
 \newcommand{\q}{{\bf q}}
 \newcommand{\po}{\stackrel{o}{\p}}
 \newcommand{\qo}{\stackrel{o}{\q}}
 \newcommand{\pmo}{\stackrel{\not\,{o}}{\p\smash{\lefteqn{_1}}}\phantom{_1}}
 \newcommand{\pno}{\stackrel{\not\,{o}}{\p\smash{\lefteqn{_2}}}\phantom{_1}}
 \newcommand{\poo}{\stackrel{\not\,{o}}{\p\smash{\lefteqn{_3}}}\phantom{_1}}
 \newcommand{\pso}{\stackrel{o}{\p\smash{\lefteqn{'}}}\phantom{}}
 \newcommand{\psoneo}{\stackrel{o}{\p\smash{\lefteqn{'_1}}}\phantom{}}
  \newcommand{\pstwoo}{\stackrel{o}{\p\smash{\lefteqn{'_2}}}\phantom{}}
 \newcommand{\psthro}{\stackrel{o}{\p\smash{\lefteqn{'_3}}}\phantom{}}
 \newcommand{\psio}{\stackrel{o}{\p\smash{\lefteqn{'_i}}}\phantom{_1}}
 \newcommand{\psko}{\stackrel{o}{\p\smash{\lefteqn{'_k}}}\phantom{_1}}
 \newcommand{\lam}{\lambda}
 \newcommand{\D}{D\smash{\lefteqn{^{(1/2)}}}\phantom{aa1}}
 \newcommand{\Dp}{\stackrel{+}{D\smash{\lefteqn{^{(1/2)}}}}\phantom{aa1}}
 \newcommand{\Dtss}{D\smash{\lefteqn{^{(1/2)}_{\lam'_3\lam''_3}}}\phantom{aa1}}
 \newcommand{\be}{\begin{equation}}
 \newcommand{\ben}{\begin{eqnarray}}
 \newcommand{\een}{\end{eqnarray}}
 \newcommand{\ee}{\end{equation}}
 \newcommand{\ra}{\rangle}
 \newcommand{\la}{\langle}
 \newcommand{\dsty}{\displaystyle}
\begin{center}
{\large\bf
How one can estimate relativistic contribution into nucleon observables
in the relativistic constituent quark model
}

\vspace*{3mm}
{\bf T.P.Ilichova}\\[6pt]
{\it Francisk Skaryna Gomel State University, \\
Laboratory of Particle Physics, JINR}  \\
\vspace*{3mm}
{\it Talk at the VIth International School-Seminar\\
      ``Actual Problems of High Energy Physics''\\
     August 7-16, 2001, Gomel, Belarus}
\end{center}
%\vspace*{2mm}
%=======================================================================
\begin{abstract}
A nonrelativistic decomposition for the quark energy by the ratio
of the dispersion of quark momentum squared
and the effective quark mass
is
investigated in the framework of the relativistic oscillator constituent
quark model
as bound systems of three valence quarks.
It is shown that relativistic corrections are defined by dispersion of the
squared
absolute value of the quark momentum.
The variations of the quark mass and oscillator parameter are studied in
detail both in the spectrum and in the nucleon magnetic moments.
\end{abstract}
\par
The nonrelativistic constituent quark model gives good results in the study
of the static properties of the nucleon, however, the dynamical characteristics
requare calculation of the pionic
cloud or/and including quark structure parameters.
\par
This might be an indication that a static parameters are weakly sensitive
to the quark
structure, and the quark anomal magnetic moments are small and the quarks
show no
structure in the observables of the baryons magnetic moments.
In this case a spectrum and magnetic moments of the nucleon can be used for
the study a relativistic contribution separately from the
contribution of quark structure.
The static parameters also weakly depend on a potential
type~\cite{Isgur-Karl-78,shulga}.
Therefore, choosing particular type of the potential one can in general
conclude about
the relativistic contribution into model observables and spread these conclusions
into dynamical properties in our model.
\par
We will consider a generalization~\cite{Ilichova98}
of the nucleon nonrelativistic oscillator model~\cite{Isgur-Karl-78}
based on a relativistic quasipotential equation for wave function of
the relative motion
$\varphi$~\cite{Log-Tav63,Faustov KPU}:
 $$
 \delta\left(\suml_{k=1}^3 \po_k\right)
 \left(
  \suml_{k=1}^3
 E_{\po_k} - M\right)
 \varphi(\po)=
 \int V
 (\po|\pso)
 *
 $$
 \be
 *
 \delta\left(
 \suml_{k=1}^3 \psko
 -\suml_{k=1}^3 \po_k
 \right)
 \delta\left(\suml_{k=1}^3 \psko\right)
 \varphi(\pso)
  \dvolumes.
 \label{Š"}
 \ee
Where
$M$ - nucleon mass,
$V$ - quasipotential,
$\p_k$ - quark momentum,
$\po_i$ -quark momentum in the nucleon rest frame,
$\po\equiv\po_1,\po_2,\po_3$, $\pso\equiv\psoneo,\pstwoo,\psthro$,
$d\Omega_{p_k}
=d\p_k/E_{\p_k}$ - element of volume,
$m$ - quark mass.
\par
Let us suppose that the nonrelativistic decomposition
of the quark energy
$E_{\po_i}=\sqrt{m^2+\po_i^2}$ one should perform without using the ratio
$\frac{\mid\po_i\mid}{m}$,
but using
a deviation momentum from average momentum, i.e.
the dispersion of the quark momentum squared.
For this aim we introduce the definition of the
effective quark mass
$
m_{eff}=\sqrt{m^2+\la\po_k^2\ra},\;\mbox{here}\;
\la f\ra=\int
d\Omega_{\p_1}
d\Omega_{\p_2}
|
\varphi|^2 f.
$
The quark energy is:
$
E_{\po_k}=\sqrt{m^2+\la\po_k^2\ra+\po_k^2-\la\po_k^2\ra}=
\sqrt{m_{eff}^2+\Delta_k},
\;
\Delta_k=\po_k^2-\la\po_k^2\ra.
$ \\
Let us consider the nonrelativistic decomposition of the quark energy
via ratio
$\Delta_k/m_{eff}$:
$
E_{\po_k}= m_{eff} +
\frac{\Delta_k}{2m_{eff}} -
\frac{\Delta_k^2}{8m^3_{eff}} + ... .
$
Our aim is to investigate the lowest-order relativistic corrections in
the quark energy:
$$
\delta_{E_k}\equiv
\la
\mid
E_{\po_k}-E_{\po_k}^{nonrel}
\mid
\ra=
\la
\mid
E_{\po_k} - m_{eff} - \frac{\Delta_k}{2m_{eff}}
\mid
\ra
\approx
\frac{
\la
\Delta^2_k
\ra
}{8 m^3_{eff}}
=
$$
\be
=
\frac{1}{8 m^3_{eff}}
\left[
\la
\po_k^4
\ra
+
\la
\po_k^2
\ra^2
-2
\la
\po_k^2
\ra^2
\right]=
\frac{1}{8 m^3_{eff}}
\left[
\la
\po_k^4
\ra
-
\la
\po_k^2
\ra^2
\right]=
\frac{
\sigma^2_{\po_k^2}
}{8 m^3_{eff}}
\ee
where,
$
\sigma_{\Delta_k}=\sigma_{\po_k^2}=\sqrt{\la\po_k^4\ra-\la\po^2\ra^2}
$ - dispersion of momentum squared.
\par
%Unsized
The undimension parameter can be written as
$
\delta_E=
\frac{
\suml_{k=1}^3
\delta_{E_k}}
{\Bigl\la \suml_{k=1}^3E_{\po_k}\Bigr\ra}*100\%.
$
\par
Now we can
decomposite
%expand
the quark energy in the Eq.~(\ref{Š"}):
$
E_{\po_i}=\sqrt{m_{eff}^2+\Delta}
=
m_{eff}
+
\frac{\Delta_k}{2m_{eff}}
+
W_{k}^{rel}
=
m_{eff}
+
\frac{\po^2_k}{2m_{eff}}-
\frac{\la\po^2_k\ra}{2m_{eff}}+W_{k}^{rel},
$
where
$
W_k^{rel}
$ is the relativistic contribution. Thus,
%Eq.~(\ref{Š"})
%can be rewrite as:
 $$
 \delta\left(\suml_{k=1}^3 \po_k\right)
 \left(
\frac{\po_1^2}{2m_{eff}}
+
\frac{\po_2^2}{2m_{eff}}
+
\frac{\po_3^2}{2m_{eff}}
+
\sum_k W_k^{rel}
+
3m_{eff}
-
M
+
C
\right)
 \varphi(\po)=
 $$
 \be
 =\delta\left(\suml_{k=1}^3 \po_k \right)
% \delta\left(\suml_{k=1}^3 \psko\right)
 \int
\left.
V (\po|
 \pso)
\right|_
 {\sum\psko=0}
\left.
 \varphi(\pso)
\right|_
 {\sum\psko=0}
d\Omega_{\psoneo}
d\Omega_{\pstwoo}
 \label{Š"_s_eff_m}
 \ee
Here, we may to perform the overdefinition of the potential and involve
constant
$C=-\sum_k\frac{\la\po_k^2\ra}{2m_k^{eff}}$ into  the potential $V$.
%,
%becous of in quark model potential are defined
%with accuracy under constant.
\par
Choosing the potential $V$ in the
Eq.~(\ref{Š"_s_eff_m})
as a generalization of the 3-particle nonrelativistic oscillator,
Eq.~(\ref{Š"_s_eff_m})
can be represented as
(
$k$ - harmonic oscillator parameter.
)
:
 $$
 \delta\left(\suml_{k=1}^3 \po_k\right)
 \Biggl[
\frac{\po_1^2}{2m_{eff}}
+
\frac{\po_2^2}{2m_{eff}}
+
\frac{\po_3^2}{2m_{eff}}
+
\sum_k W_k^{rel}
+
3m_{eff}
-
\tilde{M}-
 $$
 \be
-k
(
\nabla_{\po_1}^2
+
\nabla_{\po_2}^2
+
\nabla_{\po_1}
\nabla_{\po_2}
)
\Biggr]
 \varphi(\po)=0
 \label{Š"_s_eff_m_s_pot}
 \ee
\par
In order to estimate the relativistic contribution into spectrum and nucleon
magnetic moments we will use wave function in the zero-order approximation.
Alter neglecting
$
\sum_k W_k^{rel}
$
in the Eq.~(\ref{Š"_s_eff_m_s_pot})
one can solve exactly the zero-order aproximation
equation~(\ref{Š"_s_eff_m_s_pot})
in analogous as in the
nonrelativistic quark model:
$
\varphi^{osc}
=
N\exp\left(
   -\frac{\dsty \k^2}{\dsty 2\lambda^2}
   -\frac{\dsty {\k'}^2}{\dsty 2{\lambda'}^2}
     \right),
$
here $\lambda^2=
\gamma^2/2,\,
{\lambda'}^2=
2\gamma^2/3,\,
\gamma^2=m_{eff}\omega,
\omega=\sqrt{3k/m_{eff}},
\k =\frac{\dsty 1}{\dsty 2}(\po_1-\po_2),\k'=(\po_1+\po_2).$
\par
The calculation of nucleon current matrix element is described in
Ref.~\cite{Ilichova_gomel_99}.
The extraction of the magnetic moment has been performed using magnetic
form factor:
$
\mu
=G_M(0)=
\lim_{\K\rightarrow 0} G_M(\K)
$
\be
\mu
=
\lim_{\K\rightarrow 0}
\frac{\dsty i(2\pi)^{3} J_2^{\frac12\frac12}(\K,\0)
 \sqrt{2E_{\K}(M + E_\K) } }
 {\dsty e\left|\K\right|}
=\lim_{\K\rightarrow 0}
\frac{\dsty i(2\pi)^{3} J_2^{\frac12\frac12}(\K,\0)
 2M }
 {\dsty e\left|\K\right|}
\ee
here
$
 J_{\mu}^{\lam'\lam}(\K, \0) \equiv
 \la\lam'\K|J_{\mu}(0)|\lam\0\ra
$
\par
 The current matrix element
%~\cite{Ilichova_gomel_99}
is represented as
:
$$
\la \K \lam'\tau'\mid J_{\mu}\mid\0 \lam\tau\ra=
3
\int d\Omega_{p_1} d\Omega_{p_2} \varphi(\po_1,\po_2,\po'_3)
\chi^{\frac{1}{2}\lam\tau}_{\lam'_1\lam'_2\lam'_3\tau_1\tau_2\tau_3}\ast
$$
$$
\ast\Bigl[
\Dp(L_{\po_{12}},\pmo)
\D(L^{-1}_\K,\p_1)
\D(L_{\p_{12}},\p_1)
\Bigr]
_{\lam'_1\lam_1}\ast
$$
$$
\ast
\Bigl[
\Dp(L_{\po_{12}},\pno)
\D(L^{-1}_\K,\p_2)
\D(L_{\p_{12}},\p_2)
\Bigr]
_{\lam'_2\lam_2}\ast
$$
\be
\Dtss
(L^{-1}_\K,\p'_3)
\la\p'_3\lam''_3\tau_3\mid j_{\mu}^{(3)}\mid
\p_3\lam_3\tau_3\ra
\chi
^{\frac{1}{2}\lam\tau}
_{\lam_1\lam_2\lam_3\tau_1\tau_2\tau_3}
\varphi(\p_1,\p_2,\p_3),
\label{current}
\ee
where
$
\Dtss
(L^{-1}_\K,\p'_3)
$ - Wigner rotation matrix.
Here, neutron and proton differs in the isospin third
projection
$\tau=+\frac{1}{2}\;(p)$, $\tau=-\frac{1}{2}\;(n)$,
SU(6)- wave function of the nucleon $1/2^+$
%$\chi^{s \lam\tau}_{\lam_1\lam_2\lam_3\tau_1\tau_2\tau_3}$
is given by~\cite{klous}.
%:
%$
%%\be
%\chi^{\frac{1}{2}\lam\tau}=\frac{1}{\sqrt{2}}(\phi_{MS}^\tau\xi_{MS}^{\lam}
%+\phi_{MA}^\tau\xi_{MA}^{\lam})
%$\\
%%\label{chi_SU(6)}
%%\ee
\par
The current matrix element and the averages $\la f\ra$ were integrated using
the following methods:
VEGAS~\cite{vegas,numeric_recipes},
MISER~\cite{miser,numeric_recipes},
method of the
simple random sampling
.
VEGAS
and
MISER
give result faster on $20\%$ with $1\%$
the statistic accuracy in despite of
the simple random sampling
method.
\par
One should be clarified the basis for our choice of the parameters region.
The values of the parameters were used as
$\omega=0.52\,GeV,\; x=m_d/m_s=0.6,\;
\Delta m=m_s-m_d=0.28\,GeV$ in the
Isgur-Karl model~\cite{Isgur-Karl-78}.
In this situation results were weakly sensitive to
$x$ and $\Delta m.$
For $x=0.6$  $m_d\approx 0.42\,GeV,\gamma^2=0.22,\;$
for $x=0.7$  $m_d\approx 0.57\,GeV,\gamma^2=0.30.$
Choosing the model parameters so as to coordinate with these values, we assume
that
$\gamma^2\in[0.06\div 0.6]\,GeV^2.$
The values of quark mass from
Isgur-Karl model ($0.42,\,0.57\,GeV $)
correspond effective
quark mass $m_{eff}$ in our model.
We suppose quark mass as following:
$m_q\in[0.1\div 0.33]\,GeV.$

\begin{center}
{\bf Conclusions.}
\end{center}
\begin{itemize}
\item
The value of the relativistic contribution in the energy
$
\delta_E
$
is less than  $20\%$
in the whole
investigated
region;
$
\delta_E
$
is less than $6\%$
for $m_q\stackrel{>}{\sim} 0.2\,GeV$.
\item
The magnetic moments of the proton and neutron are obtained
with $2\%$ accuracy from experimental data
for following parameters:\\
$[\gamma^2(GeV^2)$,
$m_q(GeV)$,$m_{eff}(GeV)]$:
$[0.065,\; 0.22\div 0.26,\;0.38\div 0.39],\quad$
$[0.07,\; 0.22\div 0.3,\;0.40\div 0.42]$,
$[0.08,\; 0.24\div 0.3,\;0.42\div 0.44],\quad$
$[0.095,\; 0.22\div 0.32,\;0.46\div 0.48]$,
$[0.1,\; 0.24\div 0.32,\;0.47\div 0.49]$.
It should be noticed, that the region with
$m_q \stackrel{>}{\sim} 0.2\;GeV,$
corresponds the good agreement with the experimental data
%simultenuosly
both
for nucleon magnetic moments
with $2\%$
accuracy and
for relativistic contribution into quark energy
$
\delta_E
$
with $6\%$
accuracy.
That has been considered as a proof of the applicability of the nonrelativistic
expansion
via ratio
$\Delta_k/m_{eff}$.
\item
In ref.~\cite{Ilichova_gomel_99} was obtained,
that best simultaneous reproduction of the proton
electric and magnetic form factors
$G_E$
and
$G_M$
in oscillator model
can be obtained
for
$m_q=0.162\;GeV$, $\gamma^2=0.35\;GeV^2.$
For this parameters the proton magnetic moments differ from experimental
data up to $20\%$.
Therefore one can suppose that
agreement data for static and dynamical characteristics can be obtained
if we allow quarks to have structure: radius or magnetic moments.
\item
Effective quark mass is not fixed
in region where nucleon magnetic moments are reproduced.
The value of
$m_q^{eff}$
depends on
$m_q$, $\gamma^2$
and varied
up
$
0.48\,GeV$
to
$1.9\,GeV$
for
$m_q\in[0.1\div 0.32],\,\gamma^2\in[0.06\div 0.6].$
\end{itemize}

\end{document}